\documentclass[12pt]{article}

\usepackage{latexsym}
\usepackage{amssymb}
\usepackage{amsbsy}
\usepackage{amsmath}
\usepackage{cite}
\usepackage{epsfig}
\usepackage{graphics,color}
\usepackage{a4}
\numberwithin{equation}{section}

\allowdisplaybreaks

\newcommand{\bra}{\langle}
\newcommand{\ket}{\rangle}

\newcommand{\hm}{\hspace*{-0.6cm}}

\newcommand{\be}{\begin{equation}}
\newcommand{\ee}{\end{equation}}

\newcommand{\bea}{\begin{eqnarray}}
\newcommand{\eea}{\end{eqnarray}}
\newcommand{\bean}{\begin{eqnarray*}}
\newcommand{\eean}{\end{eqnarray*}}

\newcommand{\half} {\frac{1}{2}}

\newcommand{\re}{\operatorname{Re}}
\newcommand{\im}{\operatorname{Im}}

\newcommand{\LT}{\cite{Cristoforetti:2012su,Cristoforetti:2012uv,Cristoforetti:2013wha,Mukherjee:2013aga}}
\newcommand{\cJ}{{\cal J}}
\newcommand{\cK}{{\cal K}}

\begin{document}

\title{\bf\large
Lefschetz thimbles and stochastic quantisation:  \\ Complex actions in the complex plane}

\author{
 Gert Aarts\thanks{email: g.aarts@swan.ac.uk} \\ 
\mbox{} \\
{\em\normalsize Department of Physics, College of Science, Swansea University}\\
{\em\normalsize Swansea, United Kingdom} \\
}

\date{August 22, 2013}

\maketitle

\begin{abstract}
Lattice field theories with a complex action can be studied numerically by allowing a complexified configuration space to be explored. Here we compare the recently introduced formulation on a Lefschetz thimble with the result from stochastic quantisation (or complex Langevin dynamics) in the case of a simple model and contrast the distributions being sampled. We also study the role of the residual phase on the Lefschetz thimble.
\end{abstract}

\maketitle



\section{Introduction}
 \label{sec:intro}
 
 Lattice field theories with a complex action or Boltzmann weight, such as QCD at nonzero baryon chemical potential, cannot be simulated with algorithms based on importance sampling, due to the numerical sign problem \cite{deForcrand:2010ys,Aarts:2013bla}. Recently a new approach \LT\ has been introduced which relies on deforming the integration contour of the path integral into the complex plane and performing Monte Carlo simulations on so-called Lefschetz thimbles
 \cite{Witten:2010cx,Witten:2010zr}, along which the imaginary part of the action is constant.\footnote{For related work on the analytic continuation of path integrals, see e.g.\ Refs.\  \cite{Guralnik:2007rx,Ferrante:2013hg,Basar:2013eka}.} In this approach a residual sign problem remains, but it is allegedly much weaker than the sign problem in the original formulation. 
 
	 In this paper, we compare the formulation on Lefschetz thimbles with another approach in which the complex plane is explored, namely stochastic quantisation or complex Langevin dynamics  \cite{Parisi:1984cs,Klauder:1983,Damgaard:1987rr} (we refer to Refs.\ \cite{Aarts:2013bla,Aarts:2013uxa} for recent reviews). We carry out this comparison in the context of a simple model, for which the (real and positive) probability distribution sampled in the complex Langevin process has been constructed recently \cite{Aarts:2013uza}. Moreover, the problem on the Lefschetz thimble can be analysed analytically. This will allow us to contrast the manner in which the complex plane is explored in both approaches and study also the role of the residual phase.

In the next section, we briefly describe  the formulation of the Lefschetz thimble,  relying heavily on Refs.\ \LT. In Sec.\ \ref{sec:cc} we remind the reader of the model studied in Ref.\ \cite{Aarts:2013uza} and summarise the results on the distribution sampled during the complex Langevin process. We identify the Lefschetz thimble for this model in Sec.\ \ref{sec:LT} and discuss the weight and its residual phase on the thimble in some detail. A comparison between the distributions encountered in the Langevin and Lefschetz formulations is provided in Sec.\ \ref{sec:dist}.
The final section concludes.

\section{A single Lefschetz thimble}
\label{sec:single}

The formulation on Lefschetz thimbles is a generalisation of the method of steepest descent, in which the integration contour is deformed in the complex plane in such a way that the imaginary part of the action is constant. For a single thimble, the resulting phase can then be taken out of the functional integral and no longer contributes to the sign problem. A residual sign problem remains, arising from the curvature of the thimble, i.e.\ the change of integration path from along the real axis to the thimble, but it is expected that this sign problem is much milder \LT. 

We start by outlining the construction of the Lefschetz thimble for the simple case of  a single thimble $\cJ_0$ in a system with 
one degree of freedom \LT.
We consider the partition function
\be
Z = \int_{-\infty}^\infty dx\, e^{-S(x)},
\ee
and observables
\be
\label{eq:obsor}
\bra O(x)\ket = \frac{1}{Z} \int_{-\infty}^\infty dx\, e^{-S(x)} O(x),
\ee
where the action $S(x)$ is complex.
Extending the variable into the complex plane, $x\to z=x+iy$, and assuming that the weight $\exp(-S(z))$ is holomorphic, we now consider the case of a single nondegenerate critical point, which is determined by
\be
\partial_z S(z) \big|_{z=z_0}=0,
\quad\quad\quad\quad\quad
\partial^2_z S(z) \big|_{z=z_0}\neq 0.
\ee
For one degree of freedom,  the thimble $\cJ_0$ is then given by the requirement that the imaginary part of the action is constant along the thimble, i.e.
\be
  \im S_{\cJ_0} \equiv \im S(z) \big|_{z\in\cJ_0} = \mbox{cst},
\ee
and that the resulting  (one-dimensional) path passes through the critical point, $z_0\in \cJ_0$. To be more precise, the stable thimble $\cJ_0$ is given by the path determined by
\be
\label{eq:ev}
\dot z = -\overline{\partial_z S(z)},
\ee
ending at $z_0$ as the fiducial time $t\to\infty$. Here the dot denotes differentiation with respect to $t$ and the overline denotes complex conjugation.
In contrast, the unstable thimble $\cK_0$ is obtained by reversing the sign of $t$. The number of stable and unstable thimbles is equal. 


The important result  \cite{Witten:2010cx} is that observables in the original formulation (\ref{eq:obsor}) can now be expressed as
\be
\label{eq:obs}
\bra O(x)\ket = \frac{1}{Z_0}  \int_{\cJ_0} dz\, e^{-\re S(z)} O(z),
\ee
with the partition function 
\be
Z_0 =   \int_{\cJ_0} dz\, e^{-\re S(z)}.
\ee
We note that the (constant) phase $\exp(-i\im S_{\cJ_0})$ has canceled in Eq.\ (\ref{eq:obs}).

The Boltzmann weight due to the action is real along the thimble. However, there is still a residual phase arising from the curvature of  the thimble.
To see this, we parameterise the thimble $\cJ_0$ by the following path in the complex plane,
\be
\cJ_0: 
\quad\quad\quad  \left( x(s), y(s)\right), 
\quad\quad\quad -\infty<s<\infty,
\ee 
and write
\be
\int_{\cJ_0} dz  = \int_{-\infty}^\infty ds\, J(s),
\ee
where $J(s)$ is the complex Jacobian
\be
\label{eq:j}
J(s) = z'(s) = x^\prime(s) + i y'(s).
\ee
Here a prime denotes differentiation with respect to $s$. The final expressions are therefore
\be
\label{eq:OJ}
\bra O(x)\ket = \frac{1}{Z_0}  \int_{-\infty}^\infty ds\, J(s) e^{-\re S(z(s))} O(z(s)),
\ee
with the partition function 
\be
Z_0 =   \int_{-\infty}^\infty ds\, J(s) e^{-\re S(z(s))}.
\ee

In the case of more than one critical point $z_k$ and associated thimble $\cJ_k$, the sum over thimbles has to be taken and the expressions are generalised as \cite{Cristoforetti:2012su,Cristoforetti:2012uv,Cristoforetti:2013wha,Mukherjee:2013aga,Witten:2010cx}
\be
\label{eq:obs2}
\bra O(x)\ket = \frac{1}{Z}  \sum_k m_k \int_{\cJ_k} dz\, e^{-S(z)} O(z),
\ee
with the partition function 
\be
Z =  \sum_k m_k \int_{\cJ_k} dz\, e^{-S(z)}.
\ee
Here the integer coefficients $m_k$ are the intersection numbers between the original domain of integration and the unstable thimbles $\cK_k$.  The  phases $\exp(-i\im S_{\cJ_k})$ then no longer cancel in Eq.\ (\ref{eq:obs2}).

\section{Complex Langevin dynamics}
\label{sec:cc}

When stochastic quantisation is applied to theories with a complex action, the complexified configuration space is explored stochastically due to the complex drift term appearing in the Langevin equation  \cite{Parisi:1984cs,Klauder:1983}. This approach can solve the numerical sign problem, even when it is severe \cite{Aarts:2008wh}, but care has to be taken. A mathematical  justification of the approach can be found in Refs.\ \cite{arXiv:0912.3360,arXiv:1101.3270}. The application to QCD at nonzero baryon density is in progress   \cite{Aarts:2008rr,Aarts:2009dg,Seiler:2012wz,Sexty:2013ica}. 
 Further discussion and references can be found in  Refs.\ \cite{Aarts:2013bla,Aarts:2013uxa}.

A widely used toy model to understand the problem of complex actions and complex Langevin dynamics is the simple integral  \cite{Klauder:1985ks,Ambjorn:1985iw,Okamoto:1988ru,Duncan:2012tc,Aarts:2013uza}
\be
\label{eq:Z}
Z = \int_{-\infty}^\infty dx\, e^{-S(x)}, 
\quad\quad\quad\quad
 S=\half\sigma x^2+\frac{1}{4}\lambda x^4,
\ee
where the parameters in the action are complex-valued. Here we follow Ref.\ \cite{Aarts:2013uza} and take $\lambda$ real and positive, so that the integral exists without deformation, while $\sigma$ is taken complex. 
Exact results for expectation values $\bra x^n\ket$ can be obtained by differentiating the partition function,
\be
 Z =  \sqrt{\frac{4\xi}{\sigma}}e^{\xi} K_{-\frac{1}{4}}(\xi),
\ee
 with respect to $\sigma$. Here $\xi=\sigma^2/(8\lambda)$ and $K_p(\xi)$ is the modified Bessel function of the second kind.

In this approach, one starts from the Langevin equation for the holomorphic variable $z$, 
\be
\dot z = -\partial_zS(z) +\eta,
\ee
where the dot denotes differentiating with respect to the Langevin time $t$ and the (Gaussian) noise satisfies 
\be
\label{eq:noise}
\bra\eta(t)\eta(t')\ket=2\delta(t-t').
\ee
Writing $z=x+iy$, the complex Langevin equations then read
\be
\dot x = -\re\partial_z S(z)+ \eta, 
\quad\quad\quad
\quad\quad\quad
\dot y = -\im\partial_z S(z),
\ee
where we specialised to real noise.

\begin{figure}[t]
\begin{center}
\epsfig{figure=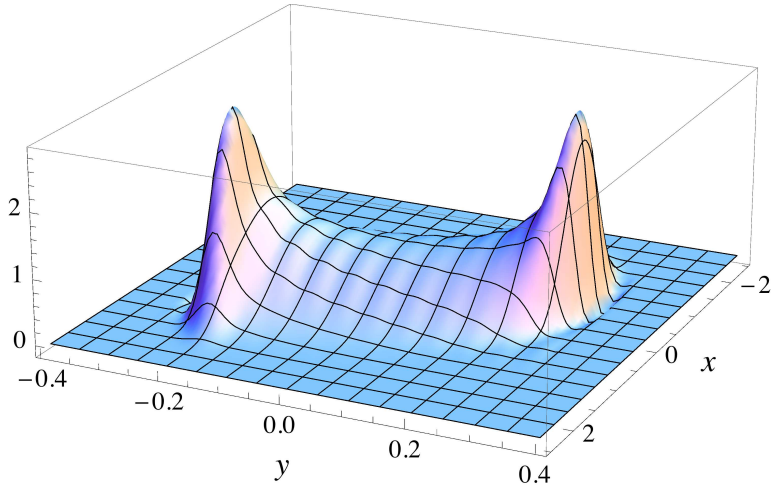 ,width=0.7\textwidth}
\end{center}
 \caption{
Distribution $P(x,y)$ in the $xy$-plane  sampled during the complex Langevin process, for $\sigma=1+i$ and $\lambda=1$. The distribution is strictly zero for $|y|>0.3029$ and drops exponentially in the $x$ direction \cite{Aarts:2013uza}. }
 \label{fig:Pxy}
\end{figure}

Expectation values are obtained by averaging over the noise.
After this averaging, holomorphic observables evolve according to
\be
\bra O\ket_{P(t)} = \int dxdy\, P(x,y;t)O(x+iy),
\ee
where the distribution $P(x,y;t)$ satisfies the Fokker-Planck equation (FPE)
\be
\label{eq:FP}
\dot P(x,y;t) = L^TP(x,y;t),
\ee
with the FP operator
\be
\label{eq:FPop}
L^T =  \partial_x \left( \partial_x + \re\partial_z S(z)  \right) + \partial_y \im\partial_z S(z).
\ee
This FPE is notoriously difficult to solve and no generic solutions are known, even for zero-dimensional integrals as in the case here.
For nontrivial solutions in specific models, see e.g.\ Refs.\ \cite{arXiv:0912.3360,Duncan:2012tc,Aarts:2013uza,Aarts:2010gr}.
In Ref.\ \cite{Aarts:2013uza}, the model under consideration was analysed in detail and the FPE was solved numerically by expanding the distribution in a  basis of Hermite functions, following the approach of Ref.\ \cite{Duncan:2012tc}. It was found that a unique stationary solution exists, which represents the (real and positive) distribution that is effectively sampled during the Langevin evolution. An example of this distribution is given in Fig.\ \ref{fig:Pxy}, for 
$\sigma=1+i$ and $\lambda=1$.
In Ref.\ \cite{Aarts:2013uza} it was also shown that, when $A>0$,  the complex Langevin process reproduces the exact results provided that 
\be
B^2<3A^2, \quad\quad\quad\quad \sigma=A+iB,
\ee
and that success and failure can be monitored by verifying the criteria for correctness  \cite{arXiv:0912.3360,arXiv:1101.3270}.
In the case of success, the distribution is strictly zero when $|y|>|y_-|$, where $y_-$ is determined by 
\be
 y_-^2 = \frac{A}{2\lambda}\left(1-\sqrt{1-\frac{B^2}{3A^2}}\right),
 \ee
 while it drops exponentially in the $x$ direction.
 In line with the mathematical foundation of the approach \cite{arXiv:0912.3360,arXiv:1101.3270}, correct results are then expected.

\section{Lefschetz thimbles for the quartic  model }
\label{sec:LT}

In this section we analyse the Lefschetz thimbles for the quartic model discussed above and study the role of the residual phase in reproducing expected results.

To construct the thimbles, we first find the critical points, determined by 
\be
\partial_zS(z) = \left(\sigma +\lambda z^2\right)z = 0, 
\quad\quad\quad 
\partial_z^2S(z) =  \sigma+3\lambda z^2\neq 0.
\ee
There are three solutions: the origin and two points in the complex plane,
\be
z_0= 0,
\quad\quad\quad\quad
z_\pm = \pm i\sqrt{\sigma/\lambda}.
\ee
Recall that $\sigma=A+iB$ is complex. The imaginary part of the action, 
\be
\im S(z) = \half B\left(x^2-y^2\right)+Axy+\lambda xy\left(x^2-y^2\right),
\ee
should be constant along a thimble. The constants are given by
\be
\im S(z_0) = 0, 
\quad\quad\quad\quad
\im S(z_\pm) = -\frac{AB}{2\lambda}.
\ee
We first discuss the points in the complex plane, $z_\pm$. Solving $\im S(z) = -AB/(2\lambda)$ yields three solutions, but it is easy to see that none of these corresponds to a thimble. There are therefore no thimbles associated with $z_\pm$.

We proceed to find the thimble associated with the origin. Solving $\im S(z) = 0$ yields again three solutions, two of which pass through $z=0$. These are the stable and unstable thimbles $\cJ_0$ and $\cK_0$. The stability can be assessed by linearising the evolution equation (\ref{eq:ev})
 around the origin, which  yields
\be
\left(  \begin{array}{c} \dot x \\ \dot y \end{array} \right)
= 
- \left(  \begin{array}{cc} A & -B \\ -B & -A \end{array} \right) \left(  \begin{array}{c} x \\  y \end{array} \right).
\ee
The matrix has eigenvalues $\pm\sqrt{A^2+B^2}$. Hence one of the solutions is stable and one is unstable, as it should be.

\begin{figure}[t]
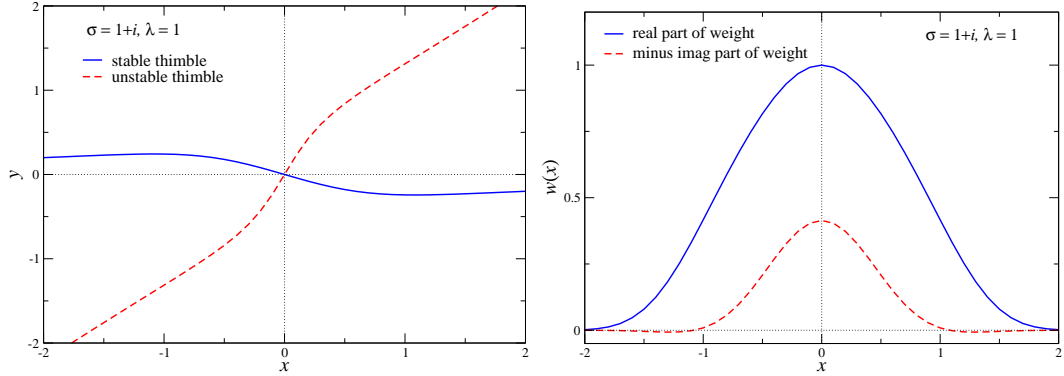

\begin{center}
\epsfig{figure=plot-thimble-st-un.eps ,width=0.48\textwidth}
\epsfig{figure=plot-thimble-weight.eps ,width=0.48\textwidth}
\end{center}
 \caption{
Stable and unstable thimble associated with the origin (left) and real and (minus the) imaginary part of the weight along the thimble, including the residual phase (right), for $\sigma=1+i$ and $\lambda=1$.}
 \label{fig:th1}
\end{figure}

The explicit expression for the stable thimble $\cJ_0$ is then (we use $x$ to parameterise the thimble)
\be
y_0(x) = \frac{1}{6\lambda x}\left( -B + e^{-i\phi}\frac{D_2}{D_1} + e^{i\phi}D_1\right),
\ee
with  $\phi=\pi/3$ and
\bea
D_1 = &&\hm  \Big(B D_3 +\sqrt{B^2 D_3^2-D_2^3}\Big)^{1/3},
\\
D_2 =&&\hm  B^2+12\lambda x^2(A+\lambda x^2),
\\
D_3 =&&\hm  B^2+18\lambda x^2(A- 2\lambda x^2).
\eea
The unstable thimble $\cK_0$ is given by the same expression, but with $\phi=-\pi/3$ (note that both expressions are real).
These thimbles are shown in Fig.\ \ref{fig:th1} (left).

The Boltzmann weight along the thimble is purely real by construction and given by
\be
w_{\rm B}(x) = \exp\left[ -S(x+iy_0(x)) \right].
\ee
However, to this the contribution from the complex Jacobian, see  Eq.\ (\ref{eq:j}), should be added, which reads
\be
J(x) = 1 + iy_0'(x).
\ee
The total weight,
\be
\label{eq:w}
w(x)  = J(x) w_{\rm B}(x),
\ee
 is therefore complex; its real and imaginary parts are shown in Fig.\ \ref{fig:th1} (right). While the real part of the weight is positive, the imaginary part  changes sign.

\begin{table}[t]
 \begin{center}
   \begin{tabular}{cclllll}
    \hline\hline
     $n$  & $\;\;\;\;\;\;$ &  $\!\!\!\!$re(Jacobian) \ \ & $\!\!\!\!$abs(Jacobian) \ \ & $\!\!\!\!$full Jacobian \ &  $\;\;$exact & Langevin \\
\hline\hline
2 & re 	& $0.207549$	 & $0.201687$	  & $0.214071$ 	& $0.214071$	& $0.2140(2)$	\\ 
   & $-$im 	& $0.111441$   & $0.108713$    & $0.0740049$ 	& $0.0740049$	& $0.0739(1)$	\\
4 & re	& $0.0912417$ & $0.0881133$ & $0.105962$		& $0.105962$	& $0.1059(2)$	\\
   & $-$im	& $0.0961076$ & $0.0929739$ & $0.070033$		& $0.070033$	& $0.0699(1)$	\\
6 & re	& $0.0712001$ & $0.0687129$ & $0.0967409$	& $0.0967409$	& $0.0967(3)$	\\
   & $-$im	& $0.124017$   & $0.119744$	  & $0.0979577$	& $0.0979577$	& $0.0978(2)$	\\
8 & re	& $0.0699149$	& $0.0675058$  & $0.118881$		& $0.118881$	& $0.1190(5)$	\\
   & $-$im	& $0.206515$   & $0.19936$ 	  & $0.17417$		& $0.17417$	& $0.1739(6)$
      \\ \hline\hline
    \end{tabular}
\end{center}
\caption{Role of the residual phase: real and (minus the) imaginary part of the observables $\bra O_n(z)\ket = \bra z^n\ket/n$ for various values of $n$, including the real part of the complex  Jacobian, the absolute value of the Jacobian and the full Jacobian. The next-to-last column is the exact result. The final column displays the complex Langevin results \cite{Aarts:2013uza}.
}
\label{tab}
\end{table}

 We can now evaluate expectation values of observables,
 \be
 O_n(z) = \frac{1}{n}z^n,
  \ee
  using Eq.\ (\ref{eq:OJ}),  where, along the thimble, $z=x+iy_0(x)$.
In Table \ref{tab} we show the results for $n=2,4,6,8$. In order to study the role of the Jacobian and the residual phase, we give a number of results.
First, we include only  the real part of $J(x)$ (which equals 1). In the next column we include the absolute value of $J(x)$, i.e.\ we ignore the residual phase, and finally we include the complete contribution. Only in the latter case, the exact results are reproduced, as expected. Interestingly, for small $n$, the first two results lie  relatively close to the exact one. However, for larger $n$ the deviation increases. We stress that the imaginary part of the weight is relevant since, even though $w_{\rm B}(x)$ is real,  the observables are complex. 
Also shown are the results from a complex Langevin simulation taken from Ref.\ \cite{Aarts:2013uza}, which are seen to agree with the exact result (within numerical error).

We conclude that when the residual phase is not incorporated correctly, exact results are not reproduced. The way in which this manifests itself may depend on the observable and is, in this model, not necessarily  small.

\section{Distributions for Langevin and Lefschetz}
\label{sec:dist}

The complex Langevin process and the Lefschetz thimble are both formulated in terms of  distributions in the complex plane. In this section we compare the two.

\begin{figure}[t]
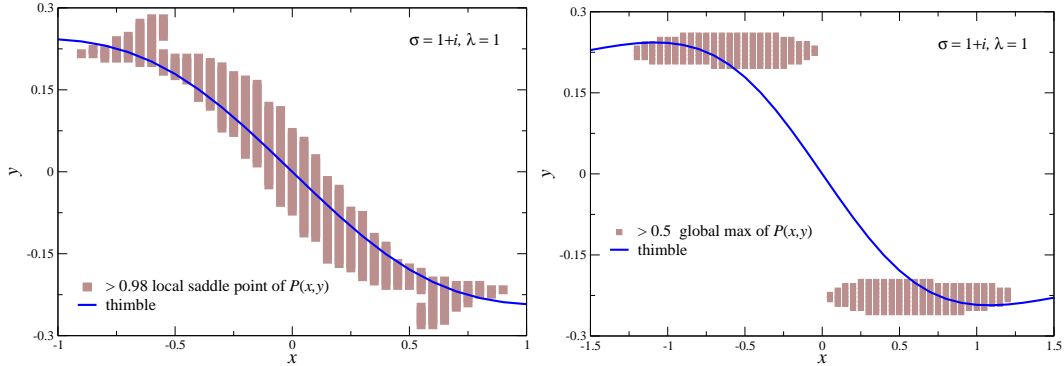

\begin{center}
\epsfig{figure=plot-thimble-CL1-v2.eps ,width=0.48\textwidth} 
\epsfig{figure=plot-thimble-CL2.eps ,width=0.48\textwidth} 
\end{center}
 \caption{
Comparison between the thimble and the distribution $P(x,y)$ sampled during the Langevin process. Left: the bars indicate the region where $P(x,y)$ is larger than $0.98$ times the value of $P(x,y)$ at the local saddle, i.e.\ the ridge in Fig.~\ref{fig:Pxy}. Right: the bars indicate the region where $P(x,y)$ is larger than 0.5 times the global maximum of $P(x,y)$, i.e.\ the two peaks in Fig.~\ref{fig:Pxy}.
}
 \label{fig:comp}
\end{figure}

We first note that the distribution sampled  in the Langevin process is a two-dimensional distribution in the complex plane, which is real and positive, i.e.\  a proper distribution. In contrast, the Lefschetz thimble is a one-dimensional path in the complex plane, on which a complex distribution is constructed. The complexity does not arise from the original weight but from the curvature of the thimble.
However, as can be gleaned from Figs.\ \ref{fig:Pxy} and \ref{fig:th1} (left), the distribution $P(x,y)$ and the thimble $\cJ_0$ are not unrelated. To demonstrate this, we compare in Fig.\ \ref{fig:comp} (left) the thimble with the region where the distribution $P(x,y)$ is larger than 0.98 times its local value at the saddle, corresponding to the ``ridge'' in Fig.~\ref{fig:Pxy}. We observe that the thimble and the ridge follow each other closely. For completeness we mention that within the numerical precision available for the construction of $P(x,y)$, the thimble and the line of saddle points of $P(x,y)$ do not seem to agree exactly.

However, the distribution of the weight between the two is quite different. Both the real and the imaginary part of the weight on the thimble peak at the origin and drop exponentially to zero as $x\to \pm\infty$, see Fig.\ \ref{fig:th1} (right). In contrast, $P(x,y)$ has two peaks close to the boundary, $y=y_-$, see Fig.\ \ref{fig:Pxy}. To highlight this difference, we show in  Fig.\ \ref{fig:comp} (right) a comparison between the thimble and the region where $P(x,y)$  is larger than 0.5 times its absolute maximum. Interestingly, we have to conclude therefore that the dominant regions contributing to the integral do not coincide, at least if we put aside the intricacies of complex phases when comparing distributions.

Finally, to estimate the importance of the residual phase, we discuss the severity of the sign problem as defined in the conventional way  \cite{deForcrand:2010ys,Aarts:2013bla}, namely by considering the expectation value of the phase factor in the theory in which the absolute value of the weight is used (the phase-quenched theory). In a theory with many degrees of freedom and a severe sign problem, the average phase factor will go to zero as the thermodynamic limit is taken. In a system with only one degree of freedom, this will not happen, but nevertheless it is interesting to compare this quantity in the original theory with the one found on the Lefschetz thimble. We write $w(x)=|w(x)|e^{i\varphi}$ and define the average phase factor in the phase-quenched theory as usual,
\be
\bra e^{i\varphi}\ket_{\rm pq} = \frac{\int dx\, w(x)}{\int dx\, |w(x)|}, 
\ee
where for the original formulation we use $w(x)=\exp(-S(x))$, while along the thimble $w(x)$ is given in Eq.\ (\ref{eq:w}).

\begin{figure}[t]
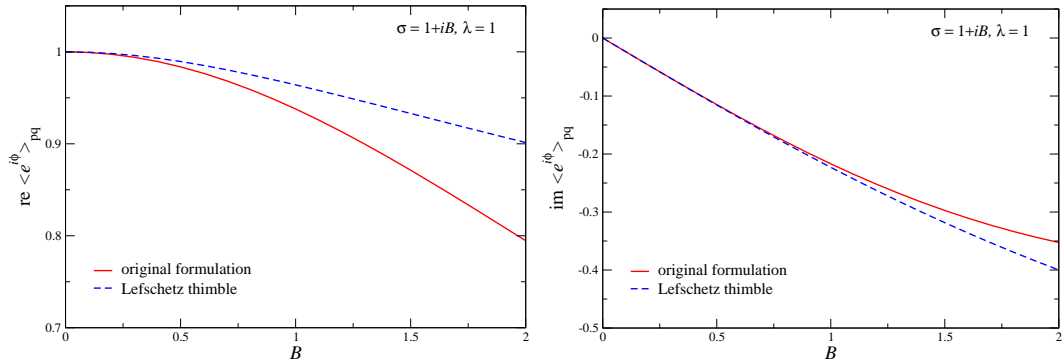

\begin{center}
\epsfig{figure=plot-phase.eps ,width=0.48\textwidth} 
\epsfig{figure=plot-phase-im.eps ,width=0.48\textwidth} 
\end{center}
 \caption{
Real (left) and imaginary (right) part of the average phase factor in the original formulation and in the formulation on the Lefschetz thimble, as a function of $B$, where $\sigma=1+iB$, with $\lambda=1$.
}
 \label{fig:phase}
\end{figure}

The results for the real and the imaginary parts of the average phase factor are shown in Fig.\ \ref{fig:phase} as a function of $B$. When $B=0$, the action is real and there is no sign problem, $\bra e^{i\varphi}\ket_{\rm pq} =1$. As $B$ is increased the real part of the average phase factor is reduced, as expected. As mentioned above, in a theory with a single degree of freedom, the sign problem is not expected to be severe. Indeed, we observe a mild sign problem, which is comparable in both formulations. There is therefore no clear indication that the sign problem due to the residual phase is considerably milder than the one in the original formulation, at least in this model and method of assessment.

\section{Conclusion}
\label{sec:conc}

In order to tackle the numerical sign problem in lattice field theory, both complex Langevin dynamics and the formulation on Lefschetz thimbles extend the original domain of integration into the complex plane. Here we compared these two approaches in the context of a simple model: interestingly we found that the two distributions that are effectively being sampled follow each other closely. 
This is one of the main results of this paper and one may wonder how general this is.
There are however important differences: during the Langevin process one encounters a two-dimensional distribution, which is real and positive, while the Lefschetz thimble is a one-dimensional path in the complex plane on which a complex distribution is constructed. The complexity, and hence a residual phase, does not arise from the original weight but from the curvature of the thimble. Moreover, we found that  the way in which the weight is distributed to be  quite different: along the thimble the maximum of the (real part of the) weight is at the origin (the critical point), while in the Langevin case, the maxima appear away from the real axis, well inside the complex plane.

As expected, the presence of the residual phase is important. Only when it is properly included are exact results reproduced. Here 
we remark that although the Boltzmann weight is real along the thimble, both the real and the imaginary part of the residual phase are relevant, since observables along the thimble are complex, even when their expectation values are real.

Finally, it would be interesting to solve the model discussed here numerically on the Lefschetz thimble, using the methods developed in Refs.\ \LT. While there is no doubt that the correct thimble will be recovered, the numerical construction of the Jacobian and its phase will be a good test for the Monte Carlo algorithm. Here we note that the correct inclusion of the residual phase appears to be more important  for observables $z^n$ with larger values of $n$.

\section*{Acknowledgments}

I thank Francesco di Renzo for the remark made in passing at Lattice 2013, which stimulated this work, 
Marco Cristoforetti and Abhishek Mukherjee for email correspondence,
Prem Kumar, Erhard Seiler, D\'enes Sexty and Nucu Stamatescu for discussion, and Pietro Giudice  and Erhard Seiler for collaboration on Ref.\ \cite{Aarts:2013uza}.
This work is supported by STFC, the Wolfson Foundation, the Royal Society and the Leverhulme Trust.

\end{document}